# Fe-DCA Metal-Organic Frameworks on the Bi$_2$Se$_3$(0001) Topological Insulator Surface

*Anna Kurowská,[1] Jakub Planer,[1] Pavel Procházka,[1] Veronika Stará,[1] Elena Vaníčková,[1] Zdeněk Endstrasser,[1] Matthias Blatnik,[1] Čestmír Drašar,[2] Jan Čechal[1,3]\**

[1] CEITEC - Central European Institute of Technology, Brno University of Technology, Purkyňova 123, 612 00 Brno, Czech Republic.

[2] University of Pardubice, Studentská 95, 53210 Pardubice, Czech Republic.

[3] Institute of Physical Engineering, Brno University of Technology, Technická 2896/2, 616 69 Brno, Czech Republic.

AUTHOR INFORMATION

**Corresponding Author**

\* E-mail: cechal@vutbr.cz (J. Č.)

**Abstract**

The formation of two-dimensional metal-organic frameworks (MOFs) on an inert surface of a topological insulator (TI) is a pathway to engineer quantum materials with exotic properties. MOFs featuring ferromagnetically coupled metal atoms are theoretically predicted to induce an exchange gap in the TI's surface band structure, potentially leading to a quantum anomalous Hall effect. However, achieving ordered MOFs on TI surfaces remains challenging due to the limited knowledge of self-assembly on these substrates. In this paper, we demonstrate self-assembly of Fe atoms and dicyanoanthracene (DCA) molecules into 2D MOFs on the $Bi_2Se_3$(0001) surface at room temperature, investigated via a combination of low-energy electron microscopy and diffraction (LEEM/LEED), scanning tunneling microscopy (STM), and ab initio calculations based on density functional theory (DFT). Two competing Fe-DCA phases form. The first phase corresponds to a close-packed $Fe_1DCA_3$ structure. In contrast, the second phase exhibits a larger unit cell with no match to either known or DFT-calculated systems, indicating a more complex bonding environment. These findings advance the understanding of the growth of MOFs on a strong topological insulator surface and provide insights for designing MOFs/TI interfaces with tailored electronic and magnetic properties.

**Introduction**

Topological insulators (TIs) belong to a class of materials having an insulating bulk and conductive surface states protected by time-reversal symmetry (TRS).[1–3] TIs were originally proposed in time-reversal-invariant systems; nonetheless, the onset of spontaneous magnetization, which breaks TRS, is expected to enable unconventional phenomena associated with nontrivial topology.[4–9] Spontaneous magnetization of TIs can emerge due to the proximity of a ferromagnetic (FM) order, such as a layer of magnetically coupled metal atoms, and open an exchange gap in the Dirac band dispersion while preserving the bulk topology and energy landscape. This could enable dissipationless charge transport, spin currents, and the emergence of Majorana fermions, with potential applications ranging from spintronics to topological quantum computation.[4,5,7]

Placing a properly designed magnetically coupled 2D layer on the TI surface thus presents a viable strategy to control the quantum states of matter and achieve new properties. In this context, on-surface coordination chemistry, where organic molecules and metal atoms self-assemble into ordered structures directly on the substrate, offers a powerful approach.[10–14] In 2D metal-organic frameworks (MOFs), organic linkers ensure the periodic order of transition metal (TM) atoms within a single layer on the surface. The local magnetic moments of TM atoms embedded in MOFs are theoretically predicted to remain unquenched, enabling the necessary exchange interaction to break TRS spontaneously.[15] Combined with flexible design, widely tunable properties, and an ability to form large-scale structures via self-assembly, MOFs are an ideal candidate to explore the interplay of strictly 2D magnetism and topological surface states, providing perspectives for realizing a robust quantum anomalous Hall and Majorana effects.[7]

The self-assembly of organic and metal-organic structures is well-studied on metal surfaces,[10,14] yet transferring this knowledge of molecular systems from metals to vdW material surfaces is often far from being straightforward, as subtle parameter changes can strongly influence the resulting structure and targeted functional properties. For non-metallic substrates, most progress has been made on epitaxial graphene,[13,16–20] followed by hBN[16,21] and NbSe$_2$[22] substrates. However, preparing a single-layer MOF on the TI surface remains a significant challenge. Here we employ a combination of low-energy electron microscopy and diffraction (LEEM/LEED), scanning tunneling microscopy (STM), and ab initio calculations based on density functional theory (DFT) to explore the formation of 2D MOFs from Fe atoms and 9,10-dicyanoantracene (DCA) molecules ([Figure 1](#)).

Among molecular and metal-organic systems studied on topological insulator substrates, metal phthalocyanines and porphyrins[23,24,33–36,25–32] have received the most attention, while other systems like TCNQ and 4F-TCNQ,[37,38] TTF,[39] alkanes,[40] fullerenes,[41] PTCDA,[42] and BDA[43] have been explored to a lesser extent. Notably, TCNQ,[37] PTCDA,[42] and BDA[43] form a highly ordered two-dimensional molecular network that weakly interacts with the substrate. In contrast, the deposition of molecules 4F-TCNQ resulted in a downward shift of the Fermi level, indicating significant charge transfer. Concerning the metal phthalocyanines (Pc), the central metal atom determines the interaction strength with the substrate and, consequently, the alignment of Pc molecules. On Bi$_2$Te$_3$, a clear trend is observed: MnPc shows strong surface interaction accompanied by the formation of an interfacial dipole, which suppresses surface molecular mobility and hinders the formation of ordered superstructures.[26,30,35] In contrast, CoPc and CuPc form well-ordered molecular superstructures, with CoPc adopting a quasi-hexagonal lattice and CuPc[26] and FePc,[24] forming square lattices with two distinct orientations, whose appearance is modulated by the extended unit cell. This trend indicates a weakening of the substrate interaction from Mn to Co, Cu, and Fe. Changing the substrate to $Bi_2Se_3$ leads to remarkable differences: while CoPc[23] still forms a quasi-hexagonal arrangement, MnPc[33] assembles in an ordered fashion, but FePc[32] and CuPc[34] do not show any ordering. This signifies the role of the substrate, as even closely related substrates can dramatically alter the adsorption properties.

In this study, we investigate the use of DCA molecules as suitable ligands (see Figure 1 for the binding motif) for the formation of MOFs on the $Bi_2Se_3$ surface. DCA has been widely reported to form MOFs with various transition metal (TM) atoms on different surfaces; some of these are predicted to exhibit magnetic properties. Recently, Lobo-Checa et al. demonstrated that $Fe_2DCA_3$ with a mixed honeycomb-kagomé (MHK) lattice on Au(111) displays ferromagnetic coupling up to 35 K,[44] making DCA a promising candidate for such studies. Additionally, progress has been made with Co-DCA on the $Bi_2Te_3$ surface, where small patches of MHK lattice form but are unstable above –20 °C.[45]

In this paper, we investigate the formation of Fe-DCA 2D MOFs on the surface of a $Bi_2Se_3$ topological insulator substrate at room temperature. Using a combination of LEEM, LEED, and STM, we observe two distinct Fe-DCA phases. The unit cell parameters, determined from the experimental data, were compared with those reported in the literature and calculated by DFT. Phase A matches the parameters of a close-packed structure, while phase B does not correspond to any previously reported or DFT calculated structures, suggesting a more complex bonding environment. Our analysis indicates that Fe-DCA does not assemble into the MHK lattice under our experimental conditions, but instead forms an alternative structure influenced by substrate templating effects and by the growth kinetics.

## Results and Discussion

Deposition of DCA molecules (Figure 1a) and Fe atoms onto a clean $Bi_2Se_3$ surface (see Supporting Information, Sections 1 and 2 for details) at room temperature results in the formation of a metal-organic structure. The LEEM bright-field image (Figure 3 in Supporting Information) shows a reduced contrast, suggesting the presence of an overlayer after the deposition, while the diffraction pattern in Figure 2a clearly shows additional spots attributed to the newly formed superstructure. The sharpness of the diffraction patterns shown in Figures 2a, c, and d indicates long-range order and reflects the quality of the grown structure. Analysis of the diffraction patterns described below reveals two distinct phases on the surface, denoted A and B. Phase B is relatively easy to obtain: at a low DCA deposition rate (0.02–0.2 ML/min), it is the only phase present for both sub-monolayer and full monolayer coverages. In contrast, phase A requires a high deposition rate of DCA and Fe, and it was observed only as a phase coexisting with phase B at or above full monolayer coverage. Upon gradual heating to 80 °C, the diffraction patterns of both phases disappear, indicating a loss of long-range order, likely due to decomposition of the metal-organic phases and desorption of DCA molecules. On a sample with coexisting phases A and B, the spots corresponding to phase A disappear at a comparatively lower temperature than those of phase B, indicating that phase B is comparatively more stable.

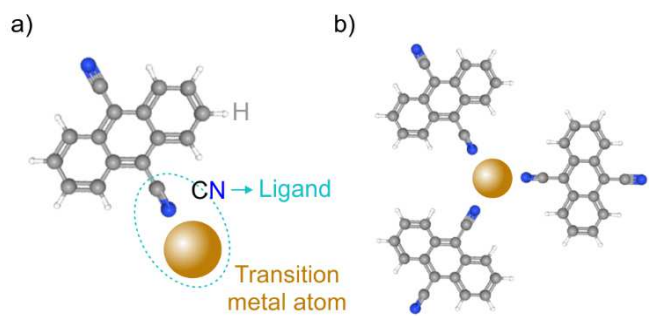

**Figure 1:** (a) Structure of DCA molecule forming a coordination bond with a transition metal atom and (b) the resulting clover-leaf motif.

The diffraction models of phases A and B shown in Figures 2b, e, and f were obtained by modeling the diffraction pattern in ProLEED Studio[46] using the measured microdiffraction patterns showing a sample area with pure phase B (Figure 2d) and the majority of phase A (Figure 2c). Lattices of phases A and B are commensurate with the substrate and are described in matrix notation as $\begin{pmatrix} 3 & -2 \\ 2 & 5 \end{pmatrix}$ and $\begin{pmatrix} 4 & -1 \\ 1 & 5 \end{pmatrix}$, respectively. The associated real-space unit cells are depicted in Figures 2g and h, respectively. The lengths of the real-space unit cell vectors (referred to as unit cell size in the following) of phases A and B are 18.1 Å and 19.0 Å, respectively, i.e., the unit cell size of phase B is 5 % larger than that of phase A. These phases have different orientations with respect to the principal $Bi_2Se_3$ substrate directions: phase A is rotated by 23.4° and phase B by 10.9° (see Figure 2e and f). In summary, phase B with a larger unit cell is more stable than phase A. From the growth and annealing behavior, we infer that phase A is a compressed phase observed only within a layer of Fe-DCA covers the entire surface.

Next, we present the STM results for the samples previously analyzed by LEEM in Figure 2, showing both structures coexisting on the surface. The STM images in Figure 3a show an array of trigonal protrusions (see detail in Figure 3b); due to their visual appearance, we refer to this motif as a cover-leaf motif. We assign the clover-leaf motif to an Fe atom coordinated with three DCA molecules, as schematically depicted in Figure 1b. These cover-leaf motifs are arranged in a hexagonal lattice, with all the motifs appearing symmetrically and oriented in the same way within a single domain. We note that the STM images were challenging to acquire due to the weak interaction of DCA with the $Bi_2Se_3$ substrate.[43] Room-temperature measurements were possible only on the sample with full monolayer coverage (presented in Figure 3); for sub-monolayer coverages, the samples had to be cooled to –100 °C to achieve atomic/molecular resolution (see Figure S4 in Supporting Information). The clover-leaf motif was the most prevalent, although at specific sample biases or under different tip conditions, other appearances were also observed, as shown in Figure S5 in the Supporting Information and in the literature.[47]

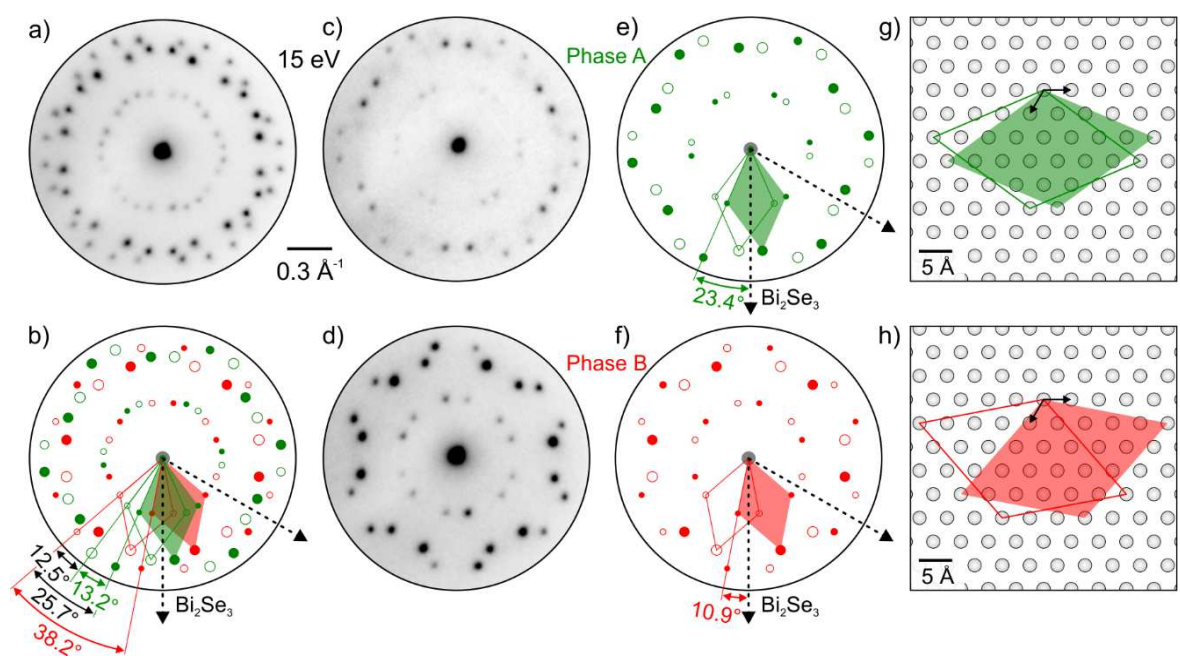

**Figure 2**: Measured and modelled diffraction patterns of Fe-DCA on the $Bi_2Se_3$(0001) surface. (a) Diffraction pattern of both phases A and B present on the surface. (b) Model of the diffraction pattern in (a). Colored arrows and parallelograms represent unit cells of phases A (green) and B (red); the gray arrows denote principal substrate directions of $Bi_2Se_3$. (c, d) Microdiffraction patterns measured on a circular area with a diameter of 3.7 μm showing the majority of phase A (c) and only phase B (d). (e, f) Diffraction models corresponding to pure phases A and B with highlighted rotation with respect to the substrate. (g, h) Real space model obtained by Fourier transform of diffraction models.

To associate the clover-leaf motif with lattices A and B, we have measured angles between the orientations of the clover-leaf motifs in distinct rotational/mirror domains, as demonstrated in Figure 4. Two distinct sets of angular differences were observed: approximately 38° corresponding to two domains of phase B (38.2°), see Figure 2b and 4a, and ~10° corresponding either to two domains of phase A (13.2°) or to an interphase of A and B (12.5°), see Figure 2b and 4b. For the latter, 5% difference in unit-cell size between the two phases is within the resolution uncertainty/error of the room-temperature STM, thus it will not allow us to distinguish the phases precisely. The results presented above suggest that both phases A and B have similar clover-leaf motifs, featuring three-fold-coordinated Fe with three DCA molecules.

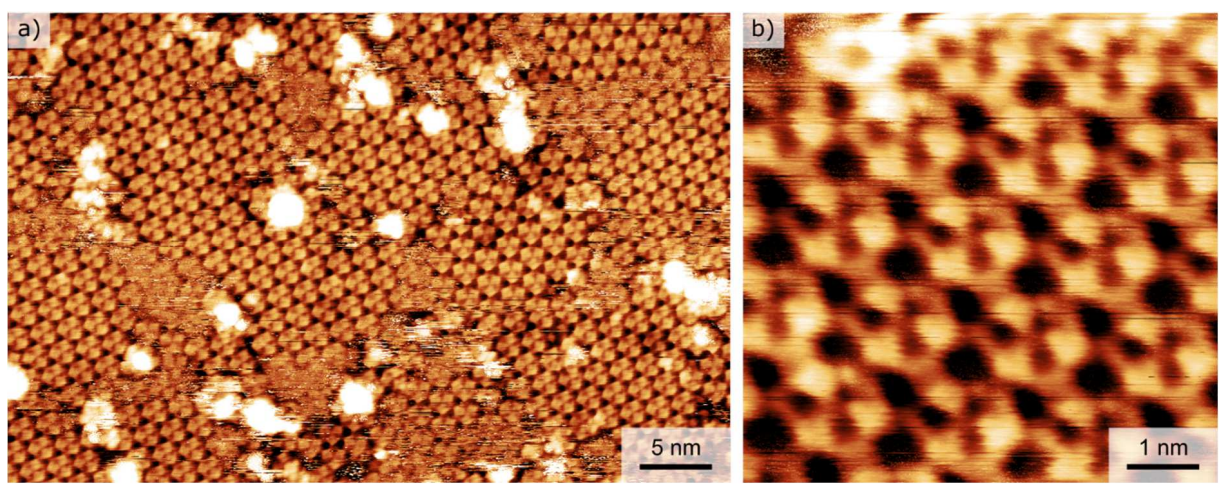

**Figure 3**: High-resolution STM images of Fe-DCA structure on $Bi_2Se_3$(0001) surface. (a) Large-scale image showing two domains of clover-leaf motifs arranged in a hexagonal lattice; the overly bright protrusions are probably clusters of Fe atoms. (b) Detailed view of the lattice of clover-leaf motifs. Scanning parameters: (a) 1.0 V, 30 pA and (b) 1.8 V, 30 pA.

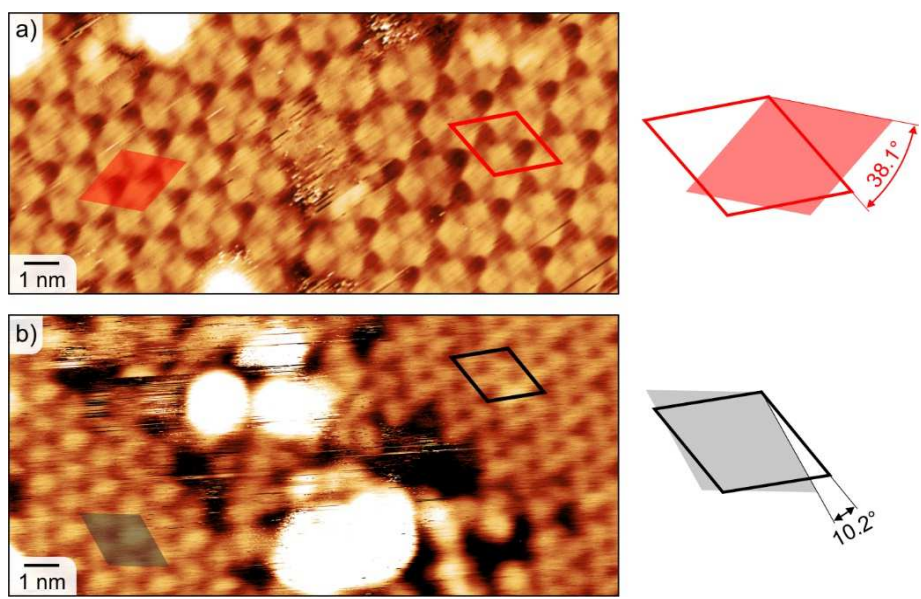

**Figure 4**: Angle analysis of Fe-DCA on $Bi_2Se_3$(0001) in STM images. (a) STM image showing two rotational domains of phase B. (b) STM image showing either two rotational domains of phase A, or the interphase of A and B. (c) Reference for angles between two symmetry-equivalent domains in the diffraction pattern. Scanning parameters: (a) 1 V, 30 pA and (b) 1.8 V, 30 pA.

To confirm the structure and composition of phases A and B, we have performed DFT calculations. There, the substrate is modeled as a single quintuple layer of $Bi_2Se_3$, and spin-orbit coupling is excluded for its high computational cost. As initial models for the calculations, we have considered the clover-leaf motif for $Fe_1DCA_3$, assuming the experimental lattices shown in Figure 5a and b. The optimized structure of $Fe_1DCA_3$ was calculated to have a unit cell vector of 17.95 Å, which closely matches the experimental value of 18.1 Å and the phase A orientation. Within this structure, the three-fold coordinated $Fe_1DCA_3$ motifs are bound together through hydrogen bonds (see Figure 5c). In contrast, phase B is larger with an experimental unit cell size of 19.0 Å, as given in Figure 5b. The total energy of the calculated gas-phase model, which matches the experimental size of the phase A, is 300 meV per

Fe$_1$DCA$_3$ formula unit higher than the gas-phase model expanded to match the unit cell of the phase B, which is caused primarily due to the weakening of hydrogen bonds between the Fe$_1$DCA$_3$ units. Furthermore, the adsorption energy of the clover phase on the Bi$_2$Se$_3$ substrate is 120 meV more favorable in the unit cell of phase A than in that of phase B, which results in a 14 % lower adsorption energy per unit area for the clover phase in the unit cell of phase A. The DFT calculations thus suggest that phase A should be the preferred clover-leaf structure, whereas the clover phase is not stable in the unit cell of phase B without an additional stabilization mechanism, which contradicts the ease of preparation of phase B. In the unit cell of phase B, there are two additional points equivalent to the position of the corner metal atoms, as highlighted in Figure 5b, which can potentially host an additional component. However, despite many efforts, such as including Fe, Bi, or Se adatoms in computations, we did not obtain any stable structure. Computational inclusion of Fe and Bi yielded the twisted MHK structure described below, whereas inclusion of Se showed low binding strength and led to a low-stability structure.

The calculated free-standing Fe$_2$DCA$_3$ with a MHK has a unit cell size of 20.6 Å, see Figure 5e. DFT optimization of Fe$_2$DCA$_3$ placed on Bi$_2$Se$_3$ results in a structure where the DCA molecules are twisted around the central atoms, suggesting that the untwisted MHK is not energetically favorable/stable on Bi$_2$Se$_3$. The twist causes shrinkage, allowing the equivalent positioning of all Fe atoms relative to the substrate atoms beneath. Such a structure would have a unit cell size of 18.96 Å, as shown in Figure 5d. DFT calculations thus show that the 2D MOF is defined by the position of the metal atom with respect to the substrate adsorption site (a position above the Se atom is preferred), which is in line with the behavior of metal phthalocyanines on Bi$_2$Se$_3$ and Bi$_2$Te$_3$ described in the introduction.

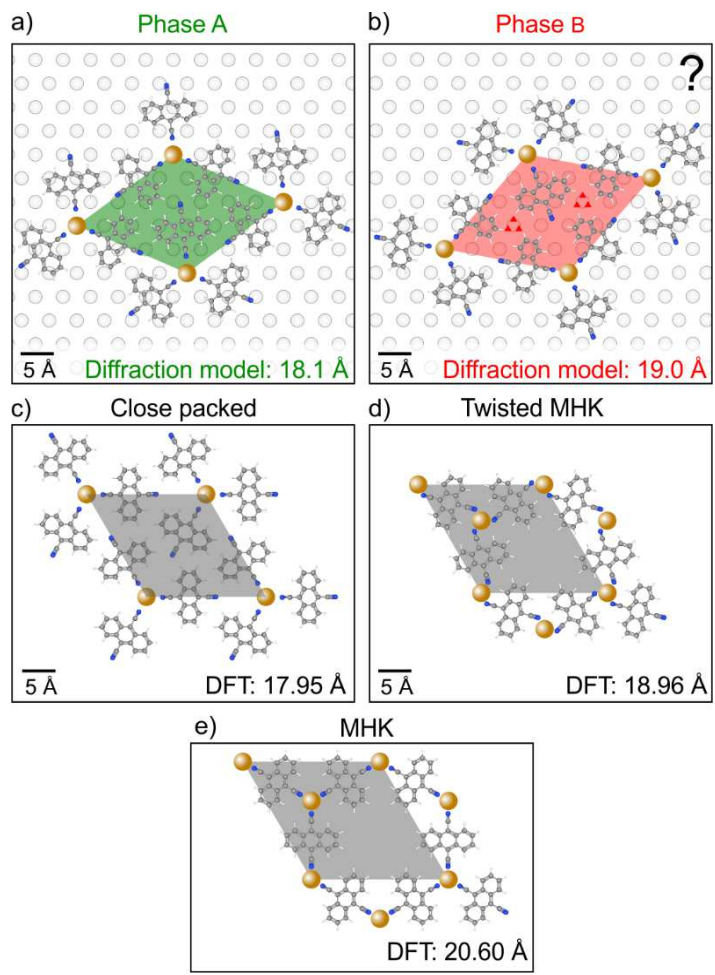

**Figure 5**: (a, b) Experimental unit cells of phases A and B determined by modelling of the diffraction patterns; the corners of the unit cells were arbitrarily positioned to the substrate lattice points. (c) Gas phase DFT model of $Fe_1DCA_3$ structure showing hydrogen-bonded clover-leaf motifs. (d) Calculated gas phase structure of $Fe_2DCA_3$ on $Bi_2Se_3$ showing twisted DCA molecules within the MHK lattice. (e) Gas phase DFT model of $Fe_2DCA_3$ with MHK lattice.

The measured real-space unit cell size of phase A is 18.1 Å, which fits well with the calculated $Fe_1DCA_3$ motif. We considered the hypothesis that phase B is $Fe_2DCA_3$ with the MHK lattice, but our experimental observations do not support this. Although the measured real-space unit cell size of phase B (19.0 Å) matches the calculated twisted MHK unit cell (18.96 Å), all STM images show phase B to be highly symmetric, which contradicts the twisted motif found in calculations. Furthermore, the MHK lattice is expected to have not only a larger unit cell size but also an interwoven structure (as seen in our results from Au(111) in Figure S6 in the Supporting Information), which is absent in the majority of our STM images, where phase B displays a clover-leaf motif. Even though HMK is sometimes imaged similarly to a clover-leaf motif, both have very distinct orientations of the motifs within the lattice as detailed in the Supporting Information, Section 6. Therefore, we reject the hypothesis that phase B is MHK. The experimental results suggest that phase B also has $Fe_1DCA_3$ clover-leaf structure, but with a larger unit cell than phase A. However, we could not find a stabilization mechanism that would computationally support its existence.

In the literature, there are two distinct motifs reported for metal (M) and DCA ($M_2DCA_3$ and $M_1DCA_3$) at a variety of substrates, i.e., Cu(111),[48–50] Ag(111),[51] Au(111),[44,52] Gr/Ir(111),[53,54] hBN,[55] NbSe$_2$[22,56]. The $M_2DCA_3$ has a fully reticulated structure with an MHK lattice, which typically forms an 8×8 superstructure with respect to the substrate; reported unit cell sizes range from 19.6 to 21.1 Å. The only exception is provided by Pawin et al., who reported $Cu_2DCA_3$ on Cu(111) with a unit cell size of 17.9 Å and a 7×7 superstructure,[49] which is likely caused by strong interactions between the molecules and the substrate. If we compare the measured phase B size of 19.0 Å with reported MHK lattices, we conclude that it is significantly smaller than expected for MHK, further supporting our conclusion that phase B is not MHK.

The second motif, a close-packed lattice of hydrogen-bonded $M_1DCA_3$ clover-leaf units, is assumed to form at higher coverages as a consequence of filling the porous $M_2DCA_3$ framework with excess DCA and subsequent transformation. It is not as prevalent in literature as MHK: $Cu_1DCA_3$ on Cu(111) is reported to have $4\sqrt{3} \times 4\sqrt{3}\ R30°$ superstructure, with unit cell size of 17.7 Å.[47,49] This number is very close to our experimental (phase A) and DFT calculated (close-packed on $Bi_2Se_3$) unit cell sizes. Conversely, $Ni_1DCA_3$ on $NbSe_2$ shows a close-packed-like structure where the whole clover-leaf motifs are twisted by a small angle within the unit cell with a size of 20.03 Å.[56] Since this is an even larger unit cell size than the phase B, the hydrogen bonding should also be significantly weakened here; however, the additional stabilization mechanism was not reported.

The $Bi_2Se_3$ substrate was recently reported to exhibit very low interaction strength with adsorbed organic molecules.[43] In contrast to graphene, which can induce the epitaxial ordering through π–π interactions, the templating effect on $Bi_2Se_3$ is much weaker. As a result, molecular layers on $Bi_2Se_3$ must be stabilized primarily by intermolecular interactions strong enough to prevent molecule detachment and desorption. However, for 2D MOFs, metal atoms may interact strongly with the $Bi_2Se_3$ substrate and occupy well-defined adsorption positions, as highlighted for metal phthalocyanines and porphyrins in the introduction. For instance, a strong affinity of Fe atoms towards Se was reported, forming FeSe complexes.[57] The strong preference for specific adsorption sites may require the metal-organic structure to be in an epitaxial relationship with the substrate to ensure that metal atoms occupy their favorable positions. Therefore, 2D MOFs in which all metal atoms are in their favorable adsorption sites are expected to be stable on the $Bi_2Se_3(0001)$ substrate.

**Conclusion**

In conclusion, we reported the synthesis of 2D metal-organic frameworks comprising Fe atoms and DCA molecules on the surface of a topological insulator, $Bi_2Se_3(0001)$. DCA coordinates with Fe atoms and forms two distinct phases, A and B, which differ in unit cell size. We have assigned phase A to be a close-packed structure, reported for metal-DCA MOFs on various substrates. In the case of phase B, neither reported nor DFT-calculated structures show a match, pointing to a more complex bonding environment. Our analysis indicates that Fe-DCA does not assemble into the MHK lattice under our experimental conditions but instead forms an alternative structure influenced by kinetic and substrate effects and atom positions. Nevertheless, it remains possible that, despite missing MHK symmetry, the close-packed structure can still exhibit magnetic coupling as it was observed on $NbSe_2$.[56] Our findings nevertheless provide insights into the formation of metal-organic structure on topological insulator surfaces, paving the way towards the synthesis of hybrid organic-inorganic material systems with the prospect of forming magnetic topological insulator via on-surface synthesis.

**Methods**

All measurements were carried out at the UHV cluster of the CEITEC Nano Core Facility. The cluster consists of a UHV transfer line and several different end stations. This allows samples to be cleaned, prepared, and characterized using multiple complementary techniques without exposure to ambient conditions. The base pressure in the UHV transfer line is $2 \times 10^{-10}$ mbar; during sample transfers (60–180 s), the pressure increases to $2 \times 10^{-9}$ mbar but quickly recovers after the movement has ceased. All chambers are pumped by standard UHV pumping equipment (turbomolecular, ion-getter, and Ti-sublimation pumps).

**Sample Preparation**

The **$Bi_2Se_3$ samples** were fabricated by heating stoichiometric 5N mixtures of Bi and Se (both from Sigma-Aldrich/Merck) with the FMC method[43] at the University of Pardubice, CZ. The resulting crystals are 4–8 mm in length, 3–6 mm in width, and up to 3 mm thick, and were mounted on a specially designed sample holder, allowing in situ exfoliation of the crystals within the UHV cluster. The process of in situ exfoliation is described in detail in our previous paper.[43] **DCA molecules are deposited** by a Createc Near-Ambient Effusion Cell from a quartz crucible at a process temperature of 55–75 °C on the sample(s) held at room temperature. The molecular powder was purchased from Merck (Sigma-Aldrich) and deposited after thorough degassing under UHV. The deposition rate was not reproducible with temperature over a longer time scale; thus, it was calibrated by depositing on an Ag(100) substrate prior to each deposition on $Bi_2Se_3$. Deposition rates ranging from 0.02 to 0.2 ML/min were tested and used. **Fe atoms are deposited** by a high-temperature cell (MBE Komponenten, WEZ) from a resistively heated quartz crucible at temperatures of 1030 °C on the sample(s) held at room temperature. The Fe pellets were purchased from Mateck and Fe was deposited after thorough UHV degassing. The deposition rate was calibrated using a quartz crystal microbalance to 0.01 pm/s; this corresponds approximately to 0.002 ML/min, while rough calculations indicate that 0.02 ML of Fe is needed for full layer coverage of the Fe-DCA MOF. In the MOF formation, the precise timing and balance of the deposition rates of both components is crucial. Due to the weak interaction between DCA and $Bi_2Se_3$ and, consequently, the low sticking coefficient, the Fe-DCA growth is performed by simultaneous deposition, usually for 10–20 minutes, with possible additional pre-deposition of only DCA and post-deposition of Fe.

**Sample Analysis**

**Scanning Tunneling Microscopy (STM)** was performed on a commercial Aarhus 150 system (SPECS) with a mounted Kolibri Sensor or a basic tungsten tip in constant-current mode at room temperature (base pressure of $2\times10^{-10}$ mbar). The corresponding STM imaging parameters are provided in the respective figure captions. **Low-Energy Electron Microscopy/Diffraction (LEEM/LEED)** images were obtained using the SPECS FE-LEEM P90 system (base pressure of $2\times10^{-10}$ mbar). Diffraction patterns are formed by collecting signals from a surface area of $15\times10$ μm$^2$. For microdiffraction analysis, this area is restricted by a mechanical aperture to a spot size of 3.7 μm. The bright-field images are obtained by the electrons from the (0,0) diffracted beam.

**Theory**

Spin-polarized Density Functional Theory calculations (DFT) were performed with the Vienna *ab initio* Simulation Package (VASP)[58] using the projector augmented wave method (PAW)[59] to treat core electrons. We used the PBE functional[60] and Grimme's pairwise D3 dispersion corrections[61] to describe the exchange-correlation energy, and a Hubbard-like Coulomb repulsion correction U-J = 4 eV in Dudarev's formulation[62] was considered for an appropriate description of Fe 3d orbitals. The energy cut-off for the plane-wave basis set was set to 500 eV. The Brillouin zone was sampled with a single Γ point. Structural optimizations were stopped when all residual forces acting on atoms in a system were smaller than 0.01 eV/Å. Following our previous work,[43] the effect of spin-orbit coupling was neglected. All interface models of $Bi_2Se_3$(0001) are composed of a single quintuple layer (QL), which showed a negligible difference in calculated surface energy (< $3\times10^{-3}$ meV/Å$^2$) with respect to the two-QL-thick slab. Additionally, a 15 Å thick vacuum layer was added in the direction perpendicular to the substrate to avoid interactions between periodically repeated replicas. All calculations account for dipole corrections to both energy and forces.

## ASSOCIATED CONTENT

Supplemental Material: (1) LEEM analysis of UHV-exfoliated $Bi_2Se_3$(0001); (2) Determination of $Bi_2Se_3$ surface termination; (3) Bright Field image of FeDCA on Bi2Se3(0001); (4) STM for submonolayer Fe-DCA coverages; (5) STM images of Fe-DCA with different tip conditions; (6) Mixed honeycomb-kagomé $Fe_2DCA_3$ on Au(111).

## AUTHOR INFORMATION

**Corresponding Authors**

* E-mail: cechal@fme.vutbr.cz (J.Č.)

**Author Contributions**

A.K., together with P.P., V.S., and M.B., prepared the samples and performed the STM and LEEM analysis. E.V. performed the LEIS experiments, Z.E. performed the LEEM-IV experiments, and both analyzed the respective data. J.P. performed DFT calculations. Č.D. $Bi_2Se_3$ prepared substrates. A.K., P.P., and J.Č. interpreted the results, and A.K. wrote the initial draft. All authors contributed to the manuscript draft and discussed the results. J.Č. initialized and supervised the work and wrote the final manuscript.

## COMPETING INTERESTS

The authors declare no competing interests.

## ACKNOWLEDGMENT

This research has been supported by GAČR, project No. 22-05114S. We acknowledge CzechNanoLab Research Infrastructure (LM2018110), supported by MEYS CR, for access to experimental facilities. The DFT computations were carried out at IT4I National Supercomputing Center supported by MEYS CR through the e-INFRA CZ, ID: 90140. M.B. was supported by the ESF under the project CZ.02.01.01/00/22_010/0002552; J.P. by the ERA Fellowship, Project No.: 101130765; and A.K. by the Brno Ph.D. Talent scholarship funded by Brno Municipality and by the Specific Research project of Brno University of Technology.

DATA AVAILABILITY

The data that support the findings of this study are available from the corresponding author upon reasonable request.

SUPPORTING INFORMATION

# Fe-DCA Metal-Organic Frameworks on Bi$_2$Se$_3$(0001) Topological Insulator Surface

*Anna Kurowská,[1] Jakub Planer,[1] Pavel Procházka,[1] Veronika Stará,[1] Elena Vaníčková,[1] Zdeněk Endstrasser,[1] Matthias Blatnik,[1] Čestmír Drašar,[2] Jan Čechal[1,3]\**

[1] CEITEC - Central European Institute of Technology, Brno University of Technology, Purkyňova 123, 612 00 Brno, Czech Republic.

[2] University of Pardubice, Studentská 95, 53210 Pardubice, Czech Republic.

[3] Institute of Physical Engineering, Brno University of Technology, Technická 2896/2, 616 69 Brno, Czech Republic.

AUTHOR INFORMATION

**Corresponding Author**

\* E-mail: cechal@vutbr.cz (J. Č.)

CONTENTS:

## 1. LEEM analysis of UHV-exfoliated $Bi_2Se_3$(0001)

A freshly exfoliated $Bi_2Se_3$ crystal was analyzed by LEEM/LEED, XPS, and STM to assess surface quality. Detailed analysis by XPS and STM is given in ref. [1]; here we focus on LEEM analysis. The exfoliated surface shows variations in real space between individual exfoliations, as detailed below. Despite the real-space variation, sharp spots in the diffraction pattern in Figure S1a reflect the three-fold symmetry of the quintuple layer and indicate a clean substrate. In real space, Figure S1b, the brighter areas are terraces (marked by green arrow) separated by step edges visualized by dark lines marked by yellow arrow; the dark circular features (blue arrow) are associated with bismuth clusters (see Section 2 in Supporting Information); their presence and number vary with each exfoliation; these clusters appear when Bi precipitates from the bulk interstitial positions to the surface.

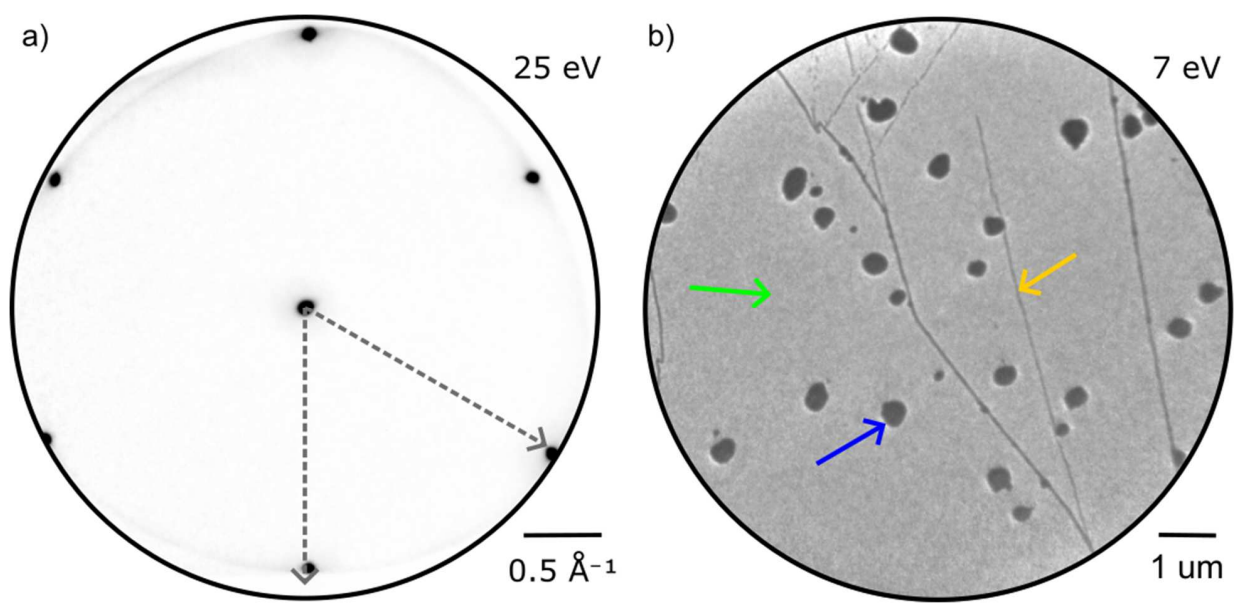

**Figure S1:** Freshly exfoliated $Bi_2Se_3$(0001) crystal characterized in LEEM/LEED. (a) LEED pattern shows clear hexagonal symmetry. (b) Bright-field LEEM image providing real real-space

view of the exfoliated surface, with highlighted features: green arrow indicates large terraces, yellow points to step edges, and blue highlights circular dark features present after exfoliation, assigned to Bi clusters precipitated from the bulk.

## 2. Determination of Bi$_2$Se$_3$ surface termination

An exfoliated surface is supposed to be Se-terminated and therefore interact weakly with adsorbates.[2] However, there has been some controversy around this topic, as a few articles reported Bi-rich termination or Bi-bilayer termination.[3–6] The appearance of this termination depends mainly on the crystalline quality (i.e., the growth process) and, to a lesser extent, on the cleaning process and sample storage conditions. To demonstrate that our substrate is indeed Se-terminated and to determine the origin of clusters observed on the exfoliated surface (Section S1), we have performed a series of experiments involving the deposition of Bi atoms onto a freshly exfoliated Bi$_2$Se$_3$ sample.

Upon the deposition of Bi, we observe growth of dark clusters in the LEEM bright field, as highlighted with a red arrow in Figure S2a and b. These clusters have a very similar appearance to those on the freshly exfoliated surface (blue arrow), suggesting a bismuth origin for the latter. We have compared LEEM-IV curves for these two clusters (Figure S2c). Because the LEEM-IV curve is highly sensitive to structure and composition, it can serve as a material and structural fingerprint.[7] Although the curves do not match in amplitude, the curves match in position and relative height of all peaks and other features. This indicates that the clusters observed after exfoliation are composed of Bi, consistent with the initial hypothesis.

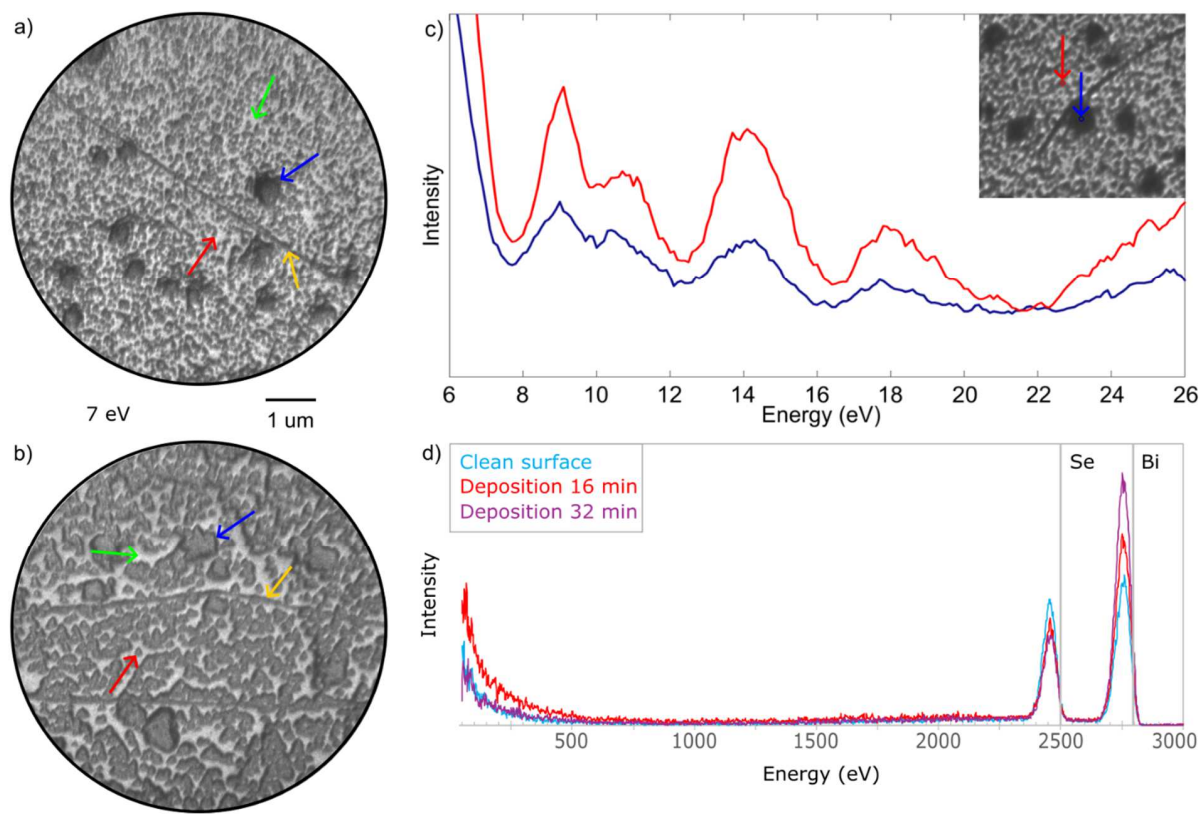

**Figure S2:** (a, b) Bright-field LEEM images taken at 7 eV on $Bi_2Se_3$(0001) after deposition of Bi for 16 min (a) and 32 min (b). The green arrow highlights terraces, the yellow points to step edges, the blue highlights circular dark features present before exfoliation, and the red indicates deposited Bi clusters. (c) LEEM-IV curves acquired on clusters present before Bi deposition (blue) and those that appeared after deposition (red) as marked in the inset; the curves are plotted for a single pixel of the image after alignment of the image series with sub-pixel precision. (d) LEIS spectra measured with He ions on a freshly exfoliated surface (Figure S1b), after 16 min (Figure S2a), and 32 min of Bi deposition (Figure S2b).

We have employed low-energy ion scattering spectroscopy (LEIS) to determine the composition of the topmost surface layer of the in-situ exfoliated $Bi_2Se_3$ samples. LEIS spectra were measured on a freshly exfoliated $Bi_2Se_3$ surface (corresponding to Figure S1), and after 16 and 32 min of Bi deposition, corresponding to Figures S2a and b, respectively. Spectra were acquired at four distinct sample locations at each stage, showing consistent results. Even on a freshly cleaved surface, we observe a significant Se peak, suggesting that Bi bilayer termination is very unlikely. With increasing Bi coverage, the Bi peak increases and the Se peak decreases in the LEIS spectra shown in Figure 2d. However, despite the high coverage of Bi clusters, there is still a quite high signal of the Se peak, likely associated with uncovered areas on the sample as marked by the green arrow in Figure S2a and b.

In the LEEM bright-field images, the lateral growth of the Bi cluster slows significantly during deposition. Accordingly, the increase in Bi LEIS intensity and the decrease in Se intensity are not proportional, as summarized in Table S1. Compared to the clean sample, the Bi signal increases by 33% after 16 min and 74% after 32 min of Bi deposition, whereas the Se peak decreases by 20% and 26%, respectively. Hence, the increase in Bi between two subsequent depositions is not accompanied by a comparable decrease in Se intensity. This points to the formation and growth of 3D clusters.

In summary, LEIS measurement suggests that Bi atoms form clusters on the $Bi_2Se_3$ surface, instead of a homogeneous bilayer, leaving uncovered Se-terminated areas. This points towards Se-termination of the freshly exfoliated surface.

**Table S1**: Comparison of Se and Bi peak intensities in LEIS spectra and their ratios for each step. Intensity values are relative to the freshly exfoliated surface. The data are treated only qualitatively by comparison with the clean substrate; a quantitative analysis was not performed due to the lack of suitable reference data for the individual elements.

| Deposition time | Se peak intensity (%) | Bi peak intenisty (%) | Se/Bi ratio |
|---|---|---|---|
| 0 min | 100 | 100 | 0.77 |
| 16 min | 80 | 133 | 0.53 |
| 32 min | 74 | 174 | 0.33 |

## 3. Bright Field image of FeDCA on Bi2Se3(0001)

There is a visible change in contrast in the LEEM bright-field image in Figure S3 after deposition compared to the clean surface. The image shows an overall reduced contrast, suggesting the presence of an overlayer, even though the individual islands cannot be resolved.

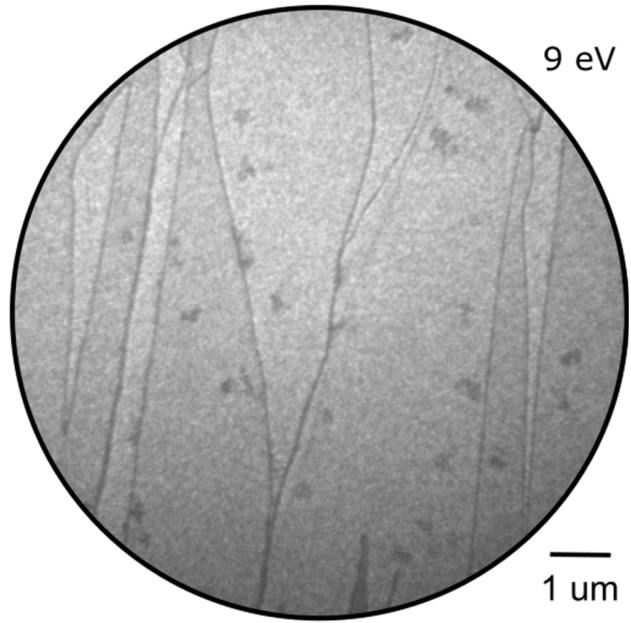

**Figure S3:** Bright-field LEEM image providing a real-space view of the $Bi_2Se_3$(0001) after deposition of Fe-DCA.

## 4. STM for submonolayer Fe-DCA coverages

While STM measurements at room temperature were very unstable for sub-monolayer Fe-DCA coverages, they still suggest the presence of a periodic arrangement, as shown in Figure S4a. After cooling to –100 °C with liquid nitrogen, molecular resolution was achieved, as shown in Figure S4b, but significant drift prevented precise length measurements.

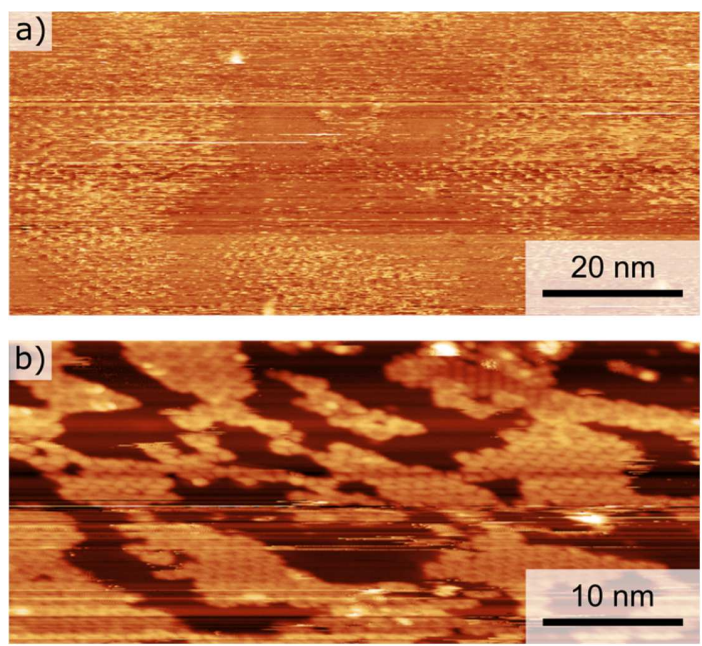

**Figure S4**: STM images measured for submonolayer coverage on a sample featuring phase B on the $Bi_2Se_3$(0001) surface. (a) Room temperature image; atomic resolution was not reached. (b) Low temperature (-100 °C) STM image. The structure is composed of clover-leaf motifs arranged in a hexagonal lattice, forming islands up to 10 nm in size. Scanning parameters: (a) 1.0 V, 30 pA and (b) 3 V, 90 pA.

## 5. STM images of Fe-DCA with different tip conditions

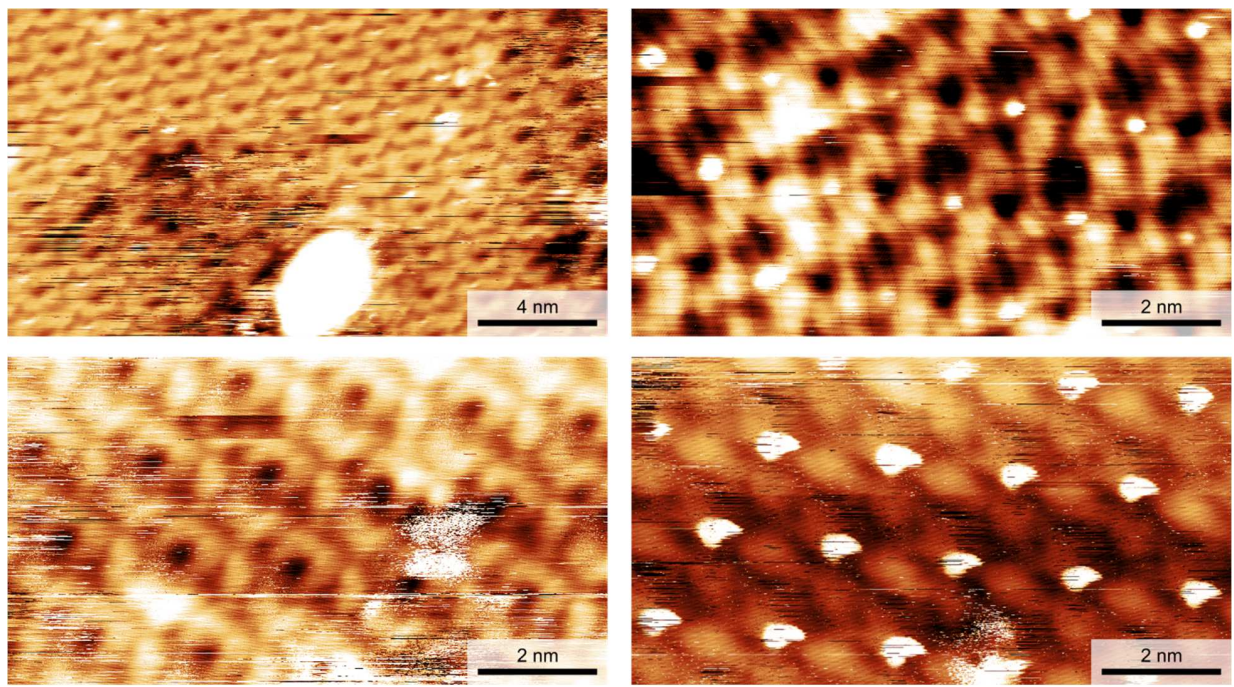

**Figure S5:** STM images showing different appearances of the clover-leaf motif with different tip conditions.

## 6. Mixed honeycomb-kagomé Fe$_2$DCA$_3$ on Au(111)

Fe$_2$DCA$_3$ with MHK structure prepared Au(111) is shown in Figure S6. At some point, the tip conditions changed, revealing a different view of the MHK lattice that resembles the clover-leaf motif. However, compared to Fe-DCA on Bi$_2$Se$_3$ in Figure S7, the angle between the individual clover motifs (marked by triangles) is different for the MHK structure on Au(111) and Phase B on Bi$_2$Se$_3$. This suggests that phase B likely does not have an MHK structure.

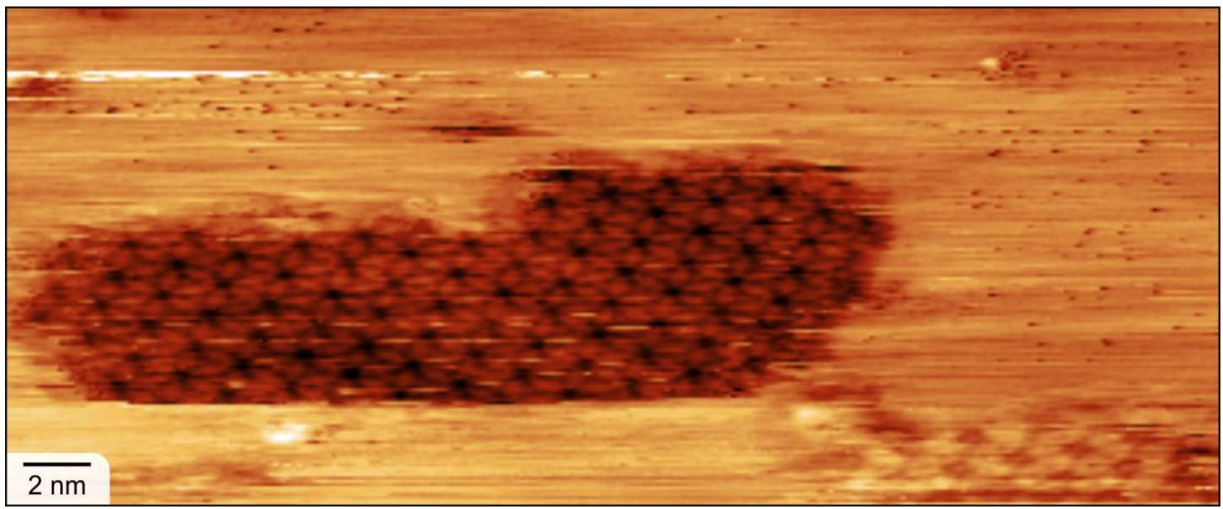

**Figure S6:** High-resolution STM images of Fe-DCA structure on Au(111) surface showing MHK structure of Fe$_2$DCA$_3$.

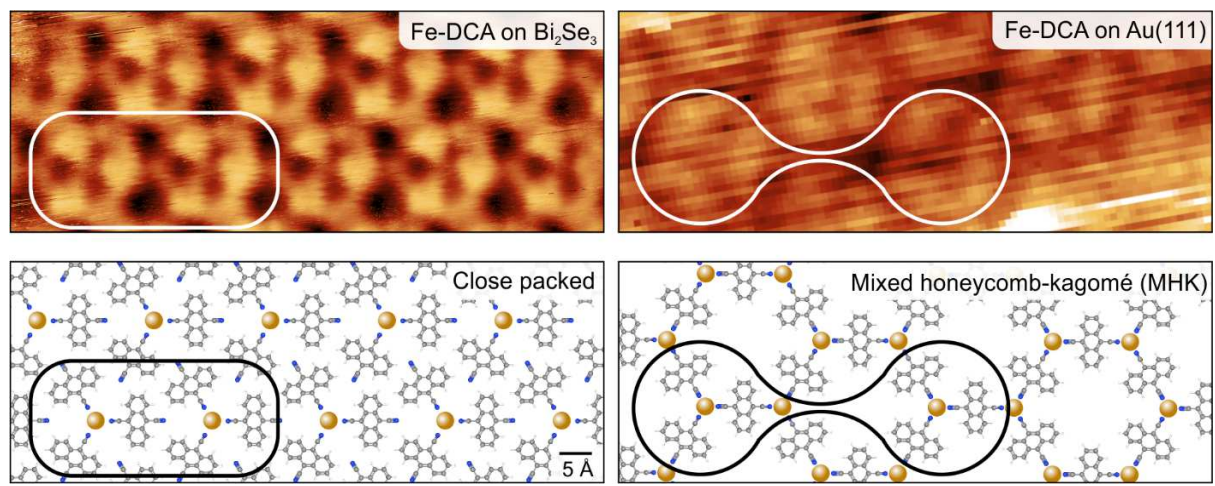

**Figure S7:** Comparison of STM images of Fe-DCA on Bi$_2$Se$_3$(0001) and Au(111). At certain tip conditions, the MHK on an Au(111) is visualized similarly to clover-like objects. These have distinct mutual orientations as compared to clover-like structures observed on Bi$_2$Se$_3$.